\documentclass[conference]{IEEEtran}
\IEEEoverridecommandlockouts
\usepackage{cite}
\usepackage{amsmath,amssymb,amsfonts}
\usepackage{algorithmic}
\usepackage{graphicx}
\usepackage{textcomp}
\usepackage{xcolor}
\usepackage[binary-units=true,per-mode=symbol]{siunitx}
\usepackage{subcaption}
\usepackage[nohyperlinks,nolist]{acronym} 
\usepackage{tikz}
\usepackage{pgfplots}
\pgfplotsset{compat=newest}
\usepgfplotslibrary{units}
\makeatletter
\pgfplotsset{
 unit code/.code 2 args=
   \begingroup
   \protected@edef\x{\endgroup\si{#2}}\x
}
\makeatother
\usetikzlibrary{pgfplots.groupplots}
\usetikzlibrary{calc}

\usepackage[absolute,showboxes]{textpos}

\pgfplotstableread[col sep = comma]{results/phases_avb.csv}\phasesavb
\pgfplotstableread[col sep = comma]{results/phases_tcp.csv}\phasestcp
\pgfplotstableread[col sep = comma]{results/recbar/ctremotesmin4.csv}\recbarmin
\pgfplotstableread[col sep = comma]{results/recbar/ctremotesmax4.csv}\recbarmax
\pgfplotstableread[col sep = comma]{results/recbar/ctremotesavg4.csv}\recbaravg
\pgfplotstableread[col sep = comma]{results/alltypes/AVB.csv}\alltypeslatencyavb
\pgfplotstableread[col sep = comma]{results/alltypes/TCP.csv}\alltypeslatencytcp
\pgfplotstableread[col sep = comma]{results/alltypes/UDP.csv}\alltypeslatencyudp
\pgfplotstableread[col sep = comma]{results/pubsubcount/1pub.csv}\pubsubcountone
\pgfplotstableread[col sep = comma]{results/pubsubcount/2pub.csv}\pubsubcounttwo
\pgfplotstableread[col sep = comma]{results/pubsubcount/3pub.csv}\pubsubcountthree
\pgfplotstableread[col sep = comma]{results/pubsubcount/4pub.csv}\pubsubcountfour
\pgfplotstableread[col sep = comma]{results/pubsubcount/5pub.csv}\pubsubcountfive
\pgfplotstableread[col sep = comma]{results/pubsubcount/6pub.csv}\pubsubcountsix
\pgfplotstableread[col sep = comma]{results/pubsubcount/7pub.csv}\pubsubcountseven
\pgfplotstableread[col sep = comma]{results/pubsubcount/8pub.csv}\pubsubcounteight
\pgfplotstableread[col sep = comma]{results/pubsubcount/9pub.csv}\pubsubcountnine
\pgfplotstableread[col sep = comma]{results/pubsubcount/10pub.csv}\pubsubcountten
\pgfplotstableread[col sep = comma]{results/pubsubcount/10pub10000.csv}\pubsubcounttenx

\usepackage{array}
\newcolumntype{L}[1]{>{\raggedright\let\newline\\\arraybackslash\hspace{0pt}}m{#1}}
\newcolumntype{C}[1]{>{\centering\let\newline\\\arraybackslash\hspace{0pt}}m{#1}}
\newcolumntype{R}[1]{>{\raggedleft\let\newline\\\arraybackslash\hspace{0pt}}m{#1}}

\definecolor{corered}{RGB}{200,0,0}
\definecolor{coreblue}{RGB}{0,46,125}
\definecolor{coregreen}{RGB}{83,129,53}

\begin{document}

\title{A QoS Aware Approach to Service-Oriented Communication in Future Automotive Networks
\thanks{This work is funded by the Federal Ministry of Education and Research of Germany (BMBF) within the SecVI project.}
}

\author{\IEEEauthorblockN{Mehmet \c{C}ak{\i}r, Timo H\"ackel,  Sandra Reider, Philipp Meyer, Franz Korf and Thomas C. Schmidt}
\IEEEauthorblockA{\textit{Dept. Computer Science},
\textit{Hamburg University of Applied Sciences}, Germany \\
{\{mehmet.cakir, timo.haeckel, sandra.reider, philipp.meyer, franz.korf, t.schmidt}\}@haw-hamburg.de}
}

\maketitle
\setlength{\TPHorizModule}{\paperwidth}
\setlength{\TPVertModule}{\paperheight}
\TPMargin{5pt}
\begin{textblock}{0.8}(0.1,0.02)
     \noindent
     \footnotesize
     If you cite this paper, please use the original reference:
     M. \c{C}ak{\i}r, T. H\"ackel,  S. Reider, P. Meyer, F. Korf and T.~C. Schmidt, "A QoS Aware Approach to Service-Oriented Communication in Future Automotive Networks," in \emph{2019 IEEE Vehicular Networking Conference (VNC) (IEEE VNC 2019)}. Los Angeles, USA, Dec. 2019.
\end{textblock}


\begin{acronym}
	\acro{API}[API]{Application Programming Interface}
	\acro{AVB}[AVB]{Audio Video Bridging}
	\acro{ARP}[ARP]{Address Resolution Protocol}
	\acro{BAG}[BAG]{Bandwidth Allocation Gap}
	\acro{BE}[BE]{Best-Effort}
	\acro{CAN}[CAN]{Controller Area Network}
	\acro{CBM}[CBM]{Credit Based Metering}
	\acro{CBS}[CBS]{Credit Based Shaping}
	\acro{CSI}[CSI]{Connection Specific Information}
	\acro{CMI}[CMI]{Class Measurement Interval}
	\acro{CoRE}[CoRE]{Communication over Realtime Ethernet}
	\acro{CORBA}[CORBA]{Common Object Request Broker Architecture}
	\acro{CT}[CT]{cross traffic}
	\acro{DoS}[DoS]{Denial of Service}
	\acro{DPI}[DPI]{Deep Packet Inspection}
	\acro{DPWS}[DPWS]{Devices Profile for Web Services}
	\acro{DDS}[DDS]{Data Distribution Service}
	\acro{ECU}[ECU]{Electronic Control Unit}
	\acroplural{ECU}[ECUs]{Electronic Control Units}
	\acro{HMI}[HMI]{Human-Machine Interface}
	\acro{IEEE}[IEEE]{Institute of Electrical and Electronics Engineers}
	\acro{IoT}[IoT]{Internet of Things}
	\acro{IFG}[IFG]{Interframe Gap}
	\acro{IPS}[IPS]{IP-based Services}
	\acro{ICT}[ICT]{Information and Communication Technology}
	\acro{JRMI}[Java RMI]{Jave Remote Method Invocation}
	\acro{JMS}[JMS]{Java Message Service}
	\acro{LIN}[LIN]{Local Interconnect Network}
	\acro{LSM}[LSM]{Local Service Manager}
	\acro{LSR}[LSR]{Local Service Registry}
	\acro{MOST}[MOST]{Media Oriented System Transport}
	\acro{MOM}[MOM]{Message Oriented Middleware}
	\acro{OSI}[OSI]{Open Systems Interconnection}
	\acro{OSGI}[OSGi]{Open Services Gateway initiative}
	\acro{OEM}[OEM]{Original Equipment Manufacturer}
	\acro{RC}[RC]{Rate-Constrained}
	\acro{REST}[ReST]{Representational State Transfer}
	\acro{RTS}[RTS]{Real-Time Services}
	\acro{SEF}[SEF]{Service Endpoint Factory}
	\acro{SRP}[SRP]{Stream Reservation Protocol}
	\acro{SRTS}[SRTS]{Static Real-Time Services}
	\acro{SOA}[SOA]{Service-Oriented Architecture}
	\acro{SOAP}[SOAP]{Simple Object Access Protocol}
	\acro{SOME/IP}[SOME/IP]{Scalable service-Oriented MiddlewarE over IP}
	\acro{SOQOSMW}[SOQoSMW]{Service-Oriented \ac{QoS} MiddleWare}
	\acro{TDMA}[TDMA]{Time Division Multiple Access}
	\acro{TSN}[TSN]{Time-Sensitive Networking}
	\acro{TT}[TT]{Time-Triggered}
	\acro{TTE}[TTE]{Time-Triggered Ethernet}
	\acro{QoS}[QoS]{Quality-of-Service}
	\acro{QoSM}[QoSM]{Quality-of-Service Manager}
	\acro{QoSNP}[QoSNP]{\ac{QoS} Negotiation Protocol}
	\acro{WS}[WS]{Web Services}
	\acro{WSDL}[WSDL]{Web Service Description Language}
\end{acronym}

\begin{abstract}
\ac{SOA} is about to enter automotive networks based on the SOME/IP middleware and an Ethernet high-bandwidth communication layer.
It promises to  meet the growing demands on connectivity and flexibility for software components in modern cars.
Largely heterogeneous service requirements and time-sensitive network functions make \ac{QoS} agreements a vital building block within future automobiles.
Existing middleware solutions, however, do not allow for a dynamic selection of \ac{QoS}.

This paper presents a service-oriented middleware for \ac{QoS} aware communication in future cars.
We contribute a protocol for dynamic \ac{QoS} negotiation along with a multi-protocol stack, which supports the different communication classes as derived from a thorough requirements analysis.
We validate the feasibility of our approach in a case study and evaluate its performance in a simulation model of a realistic in-car network.
Our findings indicate that \ac{QoS} aware communication can indeed meet the requirements, while the impact of the service negotiations and setup times of the network remain acceptable provided the cross-traffic during negotiations stays below 70\% of the available bandwidth.
\end{abstract}




 \vspace{-5pt}
\section{Introduction} 
\label{sec:introduction}
Modern cars are evolving toward software-driven cyber-physical systems with ever-growing communication demands.
The subsystems of emerging vehicular architectures are increasingly interconnected to create advanced, often complex functions.
These cross-domain functions undermine previous hard boundaries between the different automotive software domains.
The integration of new components thereby becomes more difficult as dependencies and interactions are richer and harder to predict~\cite{bcksm-scict-12}.
Besides, opening up the car network to the environment (Car-to-X) requires new communication technologies such as secure and  dynamic restful services.

The communication architecture of modern vehicles has become so complicated that it tends to stop innovation rather than promote it~\cite{f-scict-11}.
According to a broad study within the automotive industry conducted by fortiss~\cite{f-scict-11}, these problems can only be solved by redesigning the automotive communication architecture.
A high bandwidth communication backbone is proposed, in which software components communicate in a service-oriented manner~\cite{bcksm-scict-12}.

Automotive Ethernet has emerged as the next high-bandwidth communication technology for in-car backbones~\cite{mk-ae-15}.
Complementary protocols such as IEEE 802.1Q \ac{TSN}~\cite{ieee8021q-18} provide \ac{QoS}-guarantees and  have proven to meet the real-time and robustness requirements of the automotive environment~\cite{mk-ae-15}.
With the inclusion of \ac{SOME/IP}~\cite{a-somip-16}, AUTOSAR paves the way for service-oriented, IP-based communication in automotive systems.
\ac{SOME/IP} focusses on implementing \ac{SOA} while still supporting small \acp{ECU}.
Currently, a mechanism is missing that merges the concepts of \ac{SOME/IP} and QoS-enhanced communication for dynamically changing communication relations.

In this work, we contribute a \acl{QoSNP}, which allows services to dynamically achieve \ac{QoS} agreements.
Based on the \ac{QoS} requirements of automotive services, a classification into four \ac{QoS} classes is derived.
A multi-protocol stack is introduced that satisfies the heterogeneous communication demands of automotive applications.
We propose a framework for a \ac{QoS} aware service-oriented middleware that is tailored to the requirements of a car.
Thereby we focus on deadlines and latency as the essential \ac{QoS} measures of real-time systems.
In a subsequent case study,  we analyze whether the automotive requirements can be met. Evaluations are performed with the help of an automotive simulation environment in OMNeT++~\cite{mkss-smcin-19}. Communication efforts caused by the middleware are quantified using realistic vehicle communication.

The remainder of this paper is structured as follows.
Section~\ref{sec:background_&_related_work} revisits background technologies  and related work.
Section~\ref{sec:classification_of_automotive_services} analyses requirements to derive a classification of automotive services into four categories.
The design of our middleware and the \ac{QoS} negotiation are presented in Section~\ref{sec:middleware_design}.
Section~\ref{sec:evaluation} discusses the case study, in which the performance of the middleware is evaluated by simulations.
Finally, Section~\ref{sec:conclusion_&_future_work} concludes with an outlook on future work.



\section{Background and Related Work} 
\label{sec:background_&_related_work}

Ethernet has been recently extended in IEEE 802.1Q-2018~\cite{ieee8021q-18} by \acf{TSN}, which defines a set of primitives to gain the ability of forwarding real-time- and cross-traffic concurrently.
To support a variety of \ac{QoS} requirements, \ac{TSN} provides several real-time traffic classes.
These can be synchronous (\ac{TDMA}) or asynchronous such as \ac{TSN}s predecessor \ac{AVB}, which we analyzed in previous work~\cite{slksh-tiice-12}.

AUTOSAR\footnote{AUTomotive Open System ARchitecture: https://www.autosar.org/} is one of the key players in the standardization of automotive communication solutions as most \acp{ECU} are based on this platform.
Gopu et al.~\cite{gkj-soace-16} emphasize that AUTOSAR takes the challenge of finding solutions between legacy support and innovation.
With the introduction of \acf{SOME/IP}~\cite{a-somip-16} AUTOSAR was augmented by service-oriented, IP-based communication in automotive systems.
The \ac{SOME/IP} protocol consists of several components that solve problems such as data transportation, serialization, and service discovery.
\ac{SOME/IP} focusses on implementing a \ac{SOA} while still supporting small \acp{ECU}.
SOME/IP already supports static \ac{QoS} with IEEE 802.1Q priorities, however, dynamic \ac{QoS} negotiations to establish real-time communication are not supported.
Even though we do not explicitly confine our approach to the \ac{SOME/IP} protocol, the developed \ac{QoS} negotiation process can be applied to the SOME/IP middleware layer service connection process.
Whenever a new client requests information at a server, the negotiation procedure presented in this work could be applied for ongoing connections.

Besides \ac{SOME/IP}, there are many other experimental {middleware} solutions in the automotive domain and for cyber-physical systems \cite{lvh-cfiei-11,scskk-esoan-09,hssss-oasis-10,scbbz-racea-13,glv-iland-13,wzm-sodas-14}, which mainly focus on implementing IP-based communications.
In contrast, our work addresses a middleware approach that supports heterogeneous \ac{QoS} requirements, which is indispensable for future backbone and zonal architectures in automotive networks.

Due to the substantial similarity of the sensor actuator systems of a vehicle with those of industrial plants, concepts from this field of research can also be transferred to cars.
Cucinotta et al.~\cite{cmalm-rtsoa-09} as well as Jammes et al.~\cite{js-sopia-05} introduce service orientation to automation technology.
They present similar requirements as those of in-car communication architectures.
Cucinotta et al. developed the software platform \textit{RI-MACS} with a split stack of communication protocols for different classes of services, which are provided and controlled by a \ac{QoS}-based middleware.
We follow a similar approach concerning the negotiation and the automotive-specific implementation.

Menasc\'e et al.~\cite{mrg-qosms-07} emphasize that \ac{QoS}-aware middlewares are best  to satisfy as many consumers as possible in networks of components with differentiated requirements, specifically in highly distributed service-oriented systems.
In their work, they present a protocol for negotiating performance characteristics.
Negotiations are accomplished via a central \ac{QoS}-Broker, which serves as an intermediary between the services.
A client negotiates the performance characteristics for a flow with the responsible broker of the service provider.
In contrast to this centralized approach, we distribute the broker function between the provider node and the client-side.

Abdelzaher et al.~\cite{aas-qnrts-00} apply \ac{QoS}-based service-oriented communication in the real-time system  of airplanes and focus on ensuring functionality and real-time for the key components.
In their architecture, services communicate their requirements and are then assigned shares of a statically defined resource pool.
Whenever the requested resources cannot be reserved, a process of \textit{graceful degradation} begins.
Rather than denying communication to services, some services are selected according to priorities and downgraded in \ac{QoS}.
Hence, the communication for the high-priority components can be guaranteed even in overloaded situations.
The \ac{QoS} negotiation we present could be augmented by such degradation procedure, which might be a suitable solution for future autonomous driving scenarios.

Becker et al.~\cite{blc-qosac-18} describe a runtime monitoring framework based on the \ac{SOME/IP} middleware layer.
They focus on admission control, monitoring, and adaption to enforce the prescribed behavior of software components to minimize the impact of misbehavior on other flows.
This is only possible for well known components with well defined behavior and will not be enough in dynamic scenarios.
To guarantee \ac{QoS} aware communication for real-time applications this behavior must be enforced on the network level.
By introducing \ac{QoS} agreements based on client requirements, it is possible to provide such guarantees even for dynamically changing communication relations.

Another approach to \ac{QoS} based middleware solutions is \ac{DDS} from the OMG \cite{omg-dds-15}.
Again, \ac{QoS} parameters are used to control the communication properties of endpoints.
In the network, service providers are represented by a publisher and consumers by a subscriber.
The data is passed on to the publisher via a writer and received by the subscriber via a reader.
Disconnecting the publisher from the service itself allows a writer to serve multiple publishers.
Although the publish/subscribe pattern is very suitable for communication in the car, \ac{DDS} has a significant overhead and is not compatible with existing automotive communications.
This work combines the different approaches including the \ac{QoS} based publish/subscribe approach of \ac{DDS} and makes it compatible with low-level communication such as \ac{SOME/IP} to create a service-oriented middleware solution for heterogeneous applications in the automotive domain.


\section{Requirements Analysis for\\Automotive Service Classes} 
\label{sec:classification_of_automotive_services}
Automotive services are heterogeneous with differentiated \ac{QoS} requirements for communication--- all of which must be supported by
 a future automotive middleware solution~\cite{bcksm-scict-12}.
Guided by these \ac{QoS} requirements, we divide the services into classes, which then define a specific treatment by the middleware.
An overview of the different criteria for service classification is given in Table~\ref{tab:criteriontoclass}.
This table shall help in identifying the correct QoS class for a specific software component.

\begin{table*}
    \centering
    \caption{Requirement driven classification of automotive software associated with our four \ac{QoS} classes: \acf{SRTS}, \acf{RTS}, \acf{IPS}, \acf{WS}.}
    \vspace{-1mm}
    \label{tab:criteriontoclass}
    \begin{tabular}{L{2.1cm}  C{1.1cm}
                    || L{2.8cm} | L{3.15cm} | L{2.95cm} | L{2.95cm}}

        \textbf{Criterion}
        &
        \textbf{Source}
        &
        \textbf{\acs{SRTS}}
        &
        \textbf{\acs{RTS}}
        &
        \textbf{\acs{IPS}}
        &
        \textbf{\acs{WS}}
        \\\hline
        &&&&&\\[-7pt]
        \hline


        \textbf{Software Domain}
        &
        \cite{b-mddas-08} \cite{sz-ase-10} \cite{pbks-seasr-07}
        &
        Safety Electronics,\newline Engine/Powertrain
        &
        Safety Electronics,\newline Engine/Powertrain,\newline \textit{(Multimedia/HMI)}
        &
        all domains
        &
        Multimedia/HMI,\newline Passenger/Comfort,\newline Diagnostics/Infrastructure
        \\[3pt]\hline

        \textbf{Level of Abstraction}
        &
        \cite{jp-thias-16}
        &
        Signal/Physical
        &
        Data, \newline
        Signal/Physical
        &
        all levels of abstraction
        &
        Behavior, \newline
        Knowledge, \newline
        Information
        \\[3pt]\hline

        \textbf{Realtime Constraints}
        &
        \cite{jp-thias-16}
        &
        Simple Control Loops
        &
        Vehicle Dynamics Control,\newline
        Simple Control Loops
        &
        Mission Control,\newline
        Tactical Control
        &
        Cooperative Control,\newline
        Mission Control
        \\[3pt]\hline

        \textbf{Locality Constraints}
        &
        \cite{jp-thias-16}
        &
        Aggregates,\newline
        Sensors and Actuators
        &
        Vehicle,\newline
        Aggregates,\newline
        Sensors and Actuators
        &
        Vehicle,\newline
        Aggregates,\newline
        Sensors and Actuators
        &
        Global Environment,\newline
        Local Environment,\newline
        Vehicle
        \\[3pt]\hline

        \textbf{Performance of the Environment}
        &
        \cite{sz-ase-10}
        &
        Micro-Devices
        &
        PCs,\newline
        Micro-Devices
        &
        PCs,\newline
        Micro-Devices
        &
        Cloud-Infrastructure,\newline
        PCs
        \\[3pt]
    \end{tabular}
    \vspace{-6mm}
\end{table*}

    \subsection{Criteria} 
    \label{sub:criteria}

    The \textit{Automotive Software Domain} defines one category to classify groups of similar automotive software components that is most commonly used in the literature~\cite{b-mddas-08, sz-ase-10, pbks-seasr-07}.
    According to Pretschner et al.~\cite{pbks-seasr-07}, software of the domains differs in its requirements for communication deadlines, data complexity, and communication patterns.
    The most commonly described domains and requirements are:
    \begin{itemize}
        \item \textit{Multimedia/HMI --}
        Deadline: Soft, range of $100 - 250\, ms$;
        Data Complexity: Complex with large data structures;
        Communication Pattern: Discrete events and streams, off-board interface.

        \item \textit{Passenger/Comfort --}
        Deadline: Soft, range of $100 - 250\, ms$;
        Data Complexity: Mixed;
        Communication Pattern: Discrete event processing dominates over control programs.

        \item \textit{Safety Electronics --}
        Deadline: Hard, range of $1 - 10\, ms$;
        Data Complexity: Low, single values;
        Communication Pattern: Discrete event-based, strict safety requirements.

        \item \textit{Engine/Powertrain --}
        Deadline: Hard, range of micro to milliseconds;
        Data Complexity: Low, single values;
        Communication Pattern: Control algorithms dominate over discrete event processing, strict availability requirements.

        \item \textit{Diagnostics/Infrastructure --}
        Deadline: Soft and hard;
        Data Complexity: Mixed;
        Communication Pattern: Event-based software for management of the IT systems in the vehicle.
    \end{itemize}

    Jobst et al.~\cite{jp-thias-16} name three different approaches to layering automotive services based on different criteria: \textit{Level of Abstraction}, \textit{Temporal Layers}, \textit{Spatial Layers}.
    \\
    The \textit{Level of Abstraction} describes the abstraction and responsibility associated with a service.
    It focuses on the functionality and information provided to other layers and gives insight into the nature of the data to be transferred:
    \begin{itemize}
        \item \textit{Behavior --} e.g., traffic control, swarm coordination
        \item \textit{Knowledge --} e.g., ``car \SI{25}{\meter} ahead''
        \item \textit{Information --} e.g., ``object \SI{25}{\meter} ahead'' and ``car ahead''
        \item \textit{Data --} e.g processing of individual values
        \item \textit{Signal/Physical --} e.g., raw measurements, actuator control
    \end{itemize}

    The \textit{Temporal Layer} describes the timing requirements according to the cycle time that passes until the next execution.
    The cycle time limits the maximum latency of messages as otherwise, the information may be outdated.
    The maximum latency allowed for messages in the automotive domain is approximately 10\% of their cycle time. Low cycle times of only a few milliseconds are especially susceptible to jitter and latency~\cite{slksh-bhcan-15}.
    The following describes the different temporal layers:
    \begin{itemize}
        \item \textit{Cooperative Control --} e.g., coordinating swarm behavior, cycle time: unlimited
        \item \textit{Mission Control --} e.g., computing way points, cycle time: \SI{50}{\milli\second}
        \item \textit{Tactical Control --} e.g., computing trajectories, cycle time: \SI{10}{\milli\second}
        \item \textit{Vehicle Dynamics Control --} e.g. pursuing trajectories, cycle time: \SI{5}{\milli\second}
        \item \textit{Simple Control Loops --} e.g., motor and actuator control, cycle time: \SI{1}{\milli\second}
    \end{itemize}

    The \textit{Spatial Layer} (see~\cite{jp-thias-16}) describes where (geographically) within the communication network a service is located, which provides information about the appropriate communication protocols.
    By opening the vehicular network to the Internet and introducing cloud computing approaches into the car, the location of a service becomes an important criterion:
    \begin{itemize}
        \item \textit{Global Environment --} e.g., city-wide traffic
        \item \textit{Local Environment --} e.g., a swarm of five cars
	      \item \textit{Vehicle --} e.g., ECUs in specific spatial zones
        \item \textit{Aggregates --} e.g., motor control
        \item \textit{Sensors and Actuators --} e.g., temperature sensor
    \end{itemize}

\begin{figure*}
    \center
    \includegraphics[width=1\linewidth, trim= 1.2cm 0.6cm 0.9cm 0.6cm, clip=true]{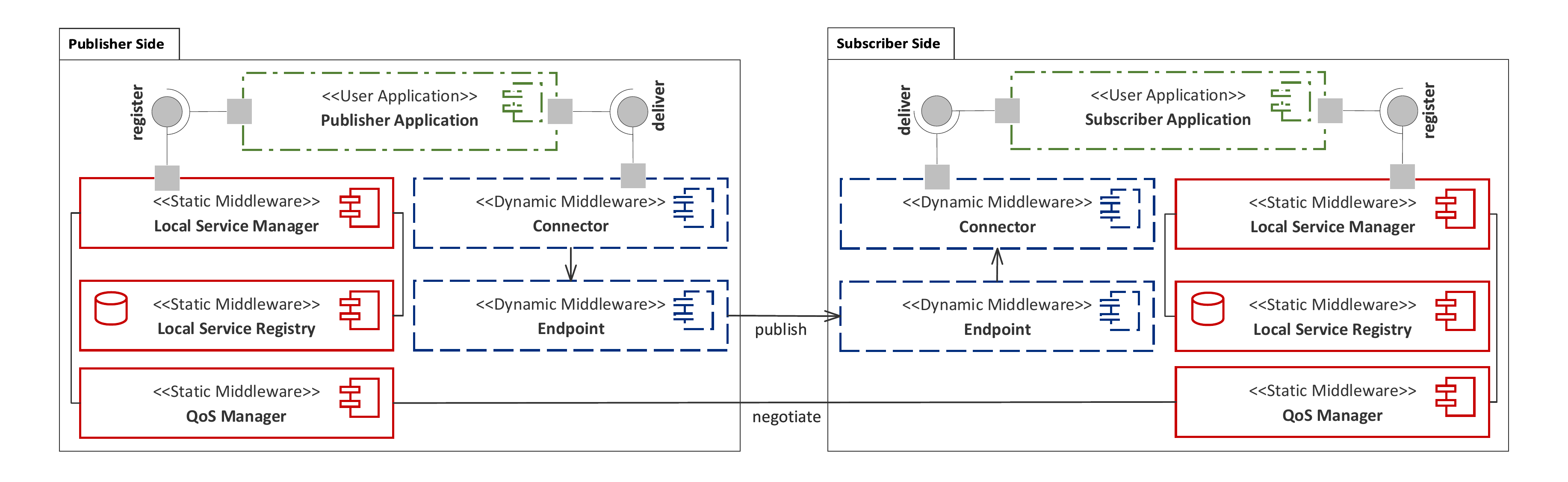}
    \caption{Middleware architecture overview containing middleware components and the communication path.}
    \label{fig:architecture}
    \vspace{-3mm}
\end{figure*}

    Sch\"auffele et al.~\cite{sz-ase-10} identify the \textit{Performance of the Environment} as another key criteria.
    This describes capabilities of the hardware components that run the service.
    The performance is divided into the following groups:
    \begin{itemize}
        \item \textit{Cloud-Infrastructure:} Scaling environments of computers with virtually unlimited performance, supporting Internet-based protocols and providing sufficient processing power for complex operations.
        \item \textit{PCs:} Computers that can host a full Linux with potent hardware.
        They support Internet-based and local protocols and provide sufficient processing power for complex operations.
        \item \textit{Micro-Devices:} Microcomputers with minimal resources.
        They support local communication interfaces but are (without hardware modules) unable to implement complex operations, such as encryption or complex serialization with reasonable effort.
    \end{itemize}


    \subsection{Classification} 
    \label{sub:classification}
   We now derive different service classes from the criteria described above, each of which represents a service requester group to the communication infrastructure.
    The characteristics and criteria may overlap in some areas, but each class focuses on  unique criteria.
    In this work, we have chosen four service classes \acl{SRTS}, \acl{RTS}, \acl{IPS}, and \acl{WS}, which are depicted in Table~\ref{tab:criteriontoclass}.

    The \textit{\acf{SRTS}} class groups various signal-based control loops that are safety-critical and highly prone to jitter and latency.
    Messages from these services need to be transmitted in a scheduled manner.
    For this reason, \ac{SRTS} services are not dynamically usable.
    Still, they can be represented as a dynamic service to the network via a gateway.

    The \textit{\acf{RTS}} class focuses on time-critical, internal communication with hard deadlines.
    \ac{RTS} are executed on PCs and Micro-Devices.
    The core domains in this class are Safety Electronics and Engine/Powertrain.
    Multimedia streams have real-time requirements as well, but they do not have the same time scale and can be softened by mechanisms such as buffering, which is impracticable for sensor data.
    For implementation in the middleware, Ethernet-based real-time transmission protocols must be used, which allow resource allocation during runtime.

    The \textit{\acf{WS}} class groups all services that focus on a high level of abstraction and global offer with no or only soft deadlines.
    Due to the global offer and the high level of abstraction, PCs with sufficient capacities and cloud infrastructure are used for these services.
    On the overall, \ac{WS} implement services of the domains Multimedia, Passenger / Comfort, and Diagnostics.
    However, the goal of all services in the \ac{WS} class is sharing information on the Internet.
    For the communication architecture, it is vital to provide these services with the usual communication standards of the Internet, e.g., authorization and serialization.
    Although the focus is on globally-used services, these services can also be used internally or in the local environment.

    The \textit{\acf{IPS}} class contains all services that do not need to be implemented in any of the other classes, meaning the focus is neither on real-time capability nor on global accessibility.
    Accordingly, services from all domains can be implemented in this class.
    To be able to communicate in the local environment, standard Internet protocols are used for data transmission.
    These services should be able to run on almost any device.



\section{Middleware Design} 
\label{sec:middleware_design}
This section presents our system architecture, which is inspired by \ac{SOME/IP}~\cite{a-somip-16} and  the architecture of \ac{DDS}~\cite{omg-dds-15}.
We adapt the protocol stack concept of Cucinotta et al.~\cite{cmalm-rtsoa-09} to the automotive domain, and design QoS negotiations similar to Menasc\'e et al.~\cite{mrg-qosms-07}.

    \subsection{System Architecture} 
    \label{sub:system_architecture}
    The middleware architecture is depicted in Figure~\ref{fig:architecture} and shows the components on the publisher and subscriber-side.
    There are three static middleware components on each host: A \acf{LSM} that manages all services on a device, a \acf{LSR} that provides a registry of all known services, and a \acf{QoSM} that implements the \ac{QoSNP} and provides brokers to carry out the negotiations.
    %
    In this work, the \ac{LSR} implements a static version, as a list of all existing services.
    In the future, this could be extended to use the \ac{SOME/IP} service discovery module.
%
    Although various communication patterns could be implemented, this work focuses on a publish/subscribe scheme since this is a very suitable pattern for in-car communication~\cite{a-somip-16}.

	A service is realized with the following three components (see Figure~\ref{fig:architecture}) to achieve a separation between data and communication; the middleware dynamically provides the latter two:
    (1)~The providing/consuming user application itself,
    (2)~an endpoint implementing the concrete protocol and transferring/receiving data to/from the network, and
    (3)~a connection module between the two.
    When an application registers a new service, the \ac{LSM} returns a connector module to the application.
	Via the connection module, the application can transmit or receive data.
    When a new connection for the service is established, the middleware connects the created endpoint to the connector module of the application.
    To create the correct connection endpoints, the \ac{QoSM} implements a \ac{QoS} negotiation described in Section~\ref{sub:quality_of_service_negotiation}.
    When the connection is set up, the data flows from the publishing application to the connector, which multiplies it for all connected endpoints.
    Each endpoint implements one of the described \ac{QoS} classes.
    The publishing endpoints then send the data to all connected subscribers in the network.
    On the subscriber-side, a connection-specific endpoint receives and unpacks the data.
    Finally, the receiving endpoint forwards the data to all subscribing applications at the host through their connector module.


    \subsection{Protocol Stack} 
    \label{sub:protocol_stack}
\begin{figure}
        \center
        \includegraphics[width=\linewidth, trim= 0.6cm 0.6cm 0.9cm 0.6cm, clip=true]{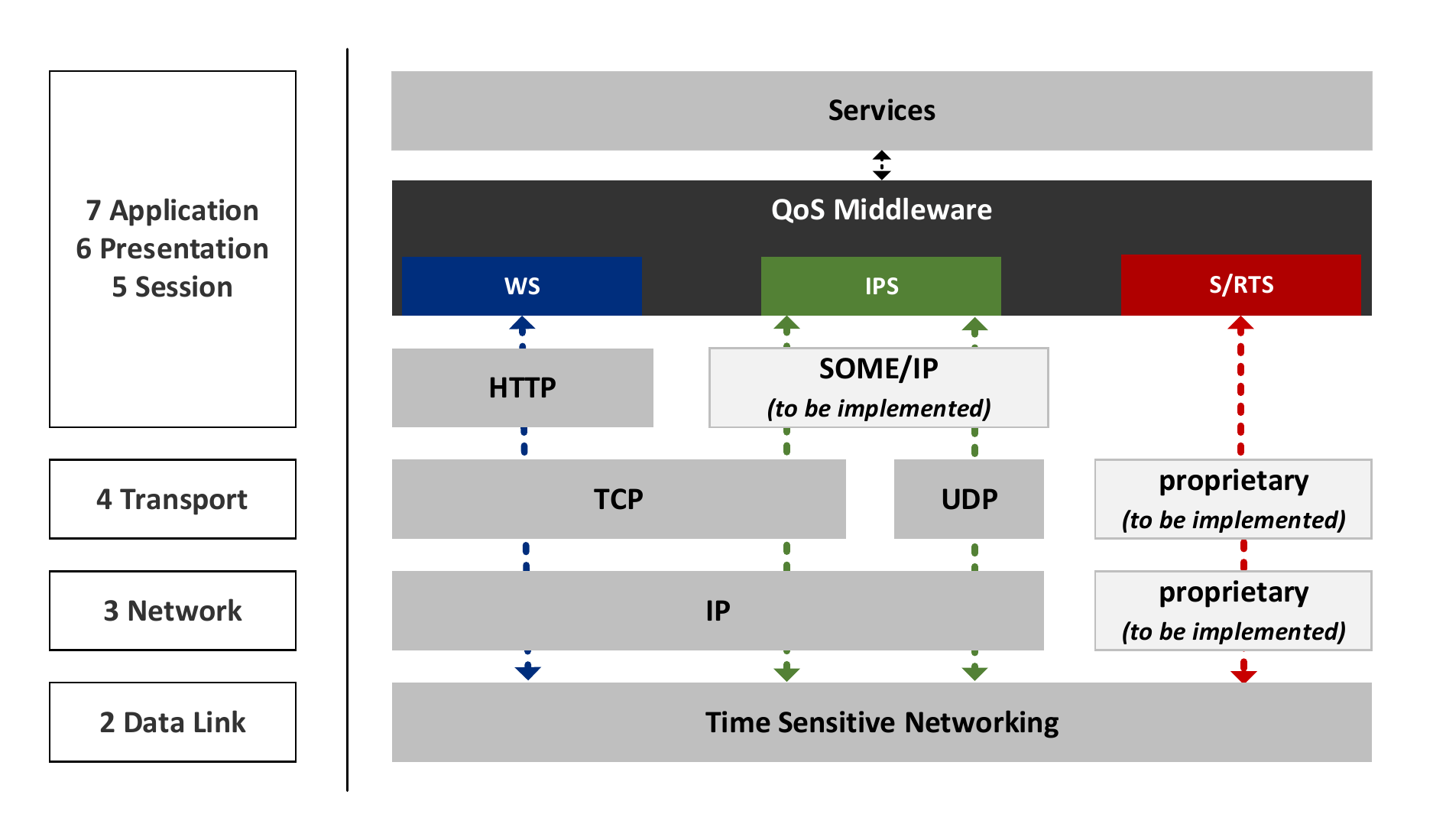}
        \caption{Multiprotocol stack according to OSI Layers divided into service classes.}
        \label{fig:protokollstack}
        \vspace{-3mm}
    \end{figure}

    Figure~\ref{fig:protokollstack} displays the protocol stack used by the middleware.
    By means of such multi-protocol stack, the middleware can select paths through the stack according to the \ac{QoS} properties.
    Appropriate endpoints can be created and connected at the application, while the specific protocol remains hidden from the application.

    The communication of the \ac{RTS} is implemented using \ac{TSN} priorities.
    By this, streams are reserved in the network, and hard real-time guarantees are provided for previously registered and reserved data volumes.
    In the current version, the \ac{RTS} class is transmitted directly onto layer 2 without using network or transport protocols.
    In a future version, UDP or \ac{SOME/IP} could be equally used provided protocols are correctly mapped.

    For services of the \ac{IPS} class, a plain IP-based stack is used that is transferred to layer 2 with best effort.
    Depending on the \ac{QoS} requirements, TCP or UDP is used at the transport level.
    In a future version, support for \ac{SOME/IP} could be added.

    The \ac{WS} class uses  HTTP on top of the TCP/IP stack.
    Thus, all modern web service implementations can be realized such as \acs{SOAP} or \acs{REST}.
    The \ac{WS} class is not  implemented yet because we focused on an existing automotive network that does not include web services.


    \subsection{Quality-of-Service Negotiation} 
    \label{sub:quality_of_service_negotiation}

    To establish communication relations with service-level policies, a \ac{QoS} agreement must be negotiated between the provider and the consumer.
    It determines the protocols that are used for transmission.
    On creation, a provider service specifies all requirements it can meet in the form of \ac{QoS} class offers.
    On the other end, the consumer specifies its request requirements.
    Since the \ac{QoS} requirements vary per client, the \ac{QoS} negotiation targets the needs of the client.

    On the client-side, the \acf{QoSM} creates a \ac{QoS} Broker that handles the negotiation for the consumer.
    If a request from a \ac{QoS} Broker arrives at the provider-side, its \ac{QoSM} first checks whether the requested service exists on this machine.
    If it exists, a \ac{QoS} Broker is also created on this side, representing the provider at the negotiation.
    Now the negotiation can begin.

\begin{figure}
        \center
        \includegraphics[width=.9\linewidth, trim= 10.9cm 28.15cm 16.5cm 15.7cm, clip=true]{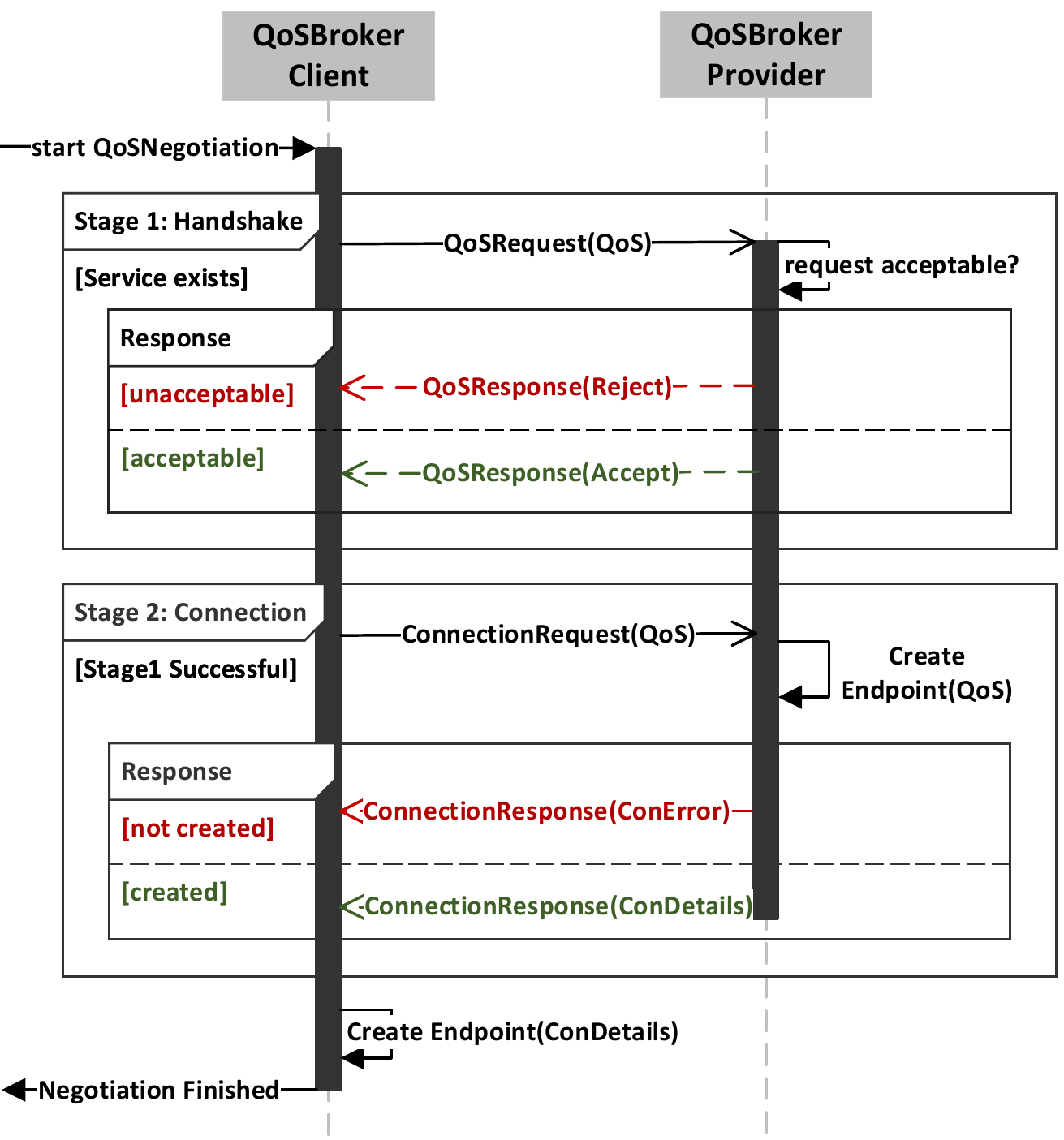}
        \caption{Sequence diagram of the connection setup with the proposed \acl{QoSNP}.}
        \label{fig:qosnpsequence}
        \vspace{-3mm}
    \end{figure}

    The \acf{QoSNP} is implemented by the \ac{QoS} Broker as a state machine, for both the consumer and the provider-side.
    Figure~\ref{fig:qosnpsequence} shows a sequence diagram of the connection setup with the \ac{QoSNP}.

    The protocol consists of two phases.
    The first phase performs a handshake.
    The consumer's broker sends a request with the required \ac{QoS} properties.
    The broker of the provider confirms that the service is reachable at the specified address and that the executing machine of the provider would generally be able to serve the consumer's request.
    The latter means that it supports both the requested protocols and QoS properties, as well as having enough resources to create a new endpoint on the system.
    If so, then a response is sent that the request is acceptable.
    If the request can not be accepted, it is also communicated.
    At this point, a future extension could be a re-negotiation with possible \textit{graceful degradation}, as described by Abdelzaher et al.~\cite{aas-qnrts-00}.
    Thus even with (partial) non-fulfillment of the requirements, data could be exchanged.

    The second phase creates the endpoints for the connection according to the agreement.
    For this purpose, a message is sent by the consumer's broker requesting the connection with the negotiated properties.
    As a result, an appropriate endpoint is created on the provider-side if it does not already exist. The middleware then connects the endpoint to the provider application via its connector module.
    The provider's broker then sends the connection details such as transport port or stream ID to the consumer's broker, and the corresponding endpoint is created.
    Finally, a connection with the endpoint of the provider is established, and the data can be sent.

    \subsection{Implementation in OMNeT++} 
    \label{sub:simulation_environment}

    The middleware has been implemented as a framework for the discrete event simulation OMNeT++\footnote{OMNeT++ Simulation Environment: https://omnetpp.org/} to make it accessible for evaluation.
    Our implementation is based on the INET framework\footnote{INET framework: https://inet.omnetpp.org/}, providing internet protocols and the CoRE simulation frameworks developed in previous work~\cite{sdks-eoifs-11}, providing real-time Ethernet protocols as well as automotive bus systems such as CAN.
    The implemented simulation model is available as open-source\footnote{\acs{SOQOSMW} simulation model: https://github.com/CoRE-RG/SOQoSMW}.


\section{Evaluation} 
\label{sec:evaluation}
We are now ready for investigating the performance of our middleware by a use case driven simulation using the OMNeT++ network simulator.
We concentrate on the real-time capabilities, which are the most critical service guaranties in the automotive domain.
Since Seyler et al.~\cite{ssgmt-fassd-15} already evaluated the impact of service discovery, we focus our evaluation on (1) the impact of the \ac{QoS} negotiation on the setup time of all services and (2) latency ranges of different \ac{QoS} classes.

Almost all \acp{ECU} of a parked vehicle are inactive before use.
As soon as the driver contacts the car, e.g., via keyless entry ECUs and services are activated step by step.
First, the services of the door ECU or the service "Active interior lighting" must be available very quickly.
Services of the engine ECU may be available later after the driver has taken a seat in his car.
In current cars the door ECU must be working after $\approx$~\SI{150}{\milli\second} to \SI{200}{\milli\second}.
Hence, negotiation times for services that must be quickly available should be at least an order of magnitude smaller.
Advantageously, only a small amount of network traffic takes place in the wake-up phase of the vehicle.

\noindent Two networks of  \SI{1}{\giga\bit\per\second} capacity will be investigated.
\begin{figure}
    \center
    \includegraphics[width=1.\linewidth]{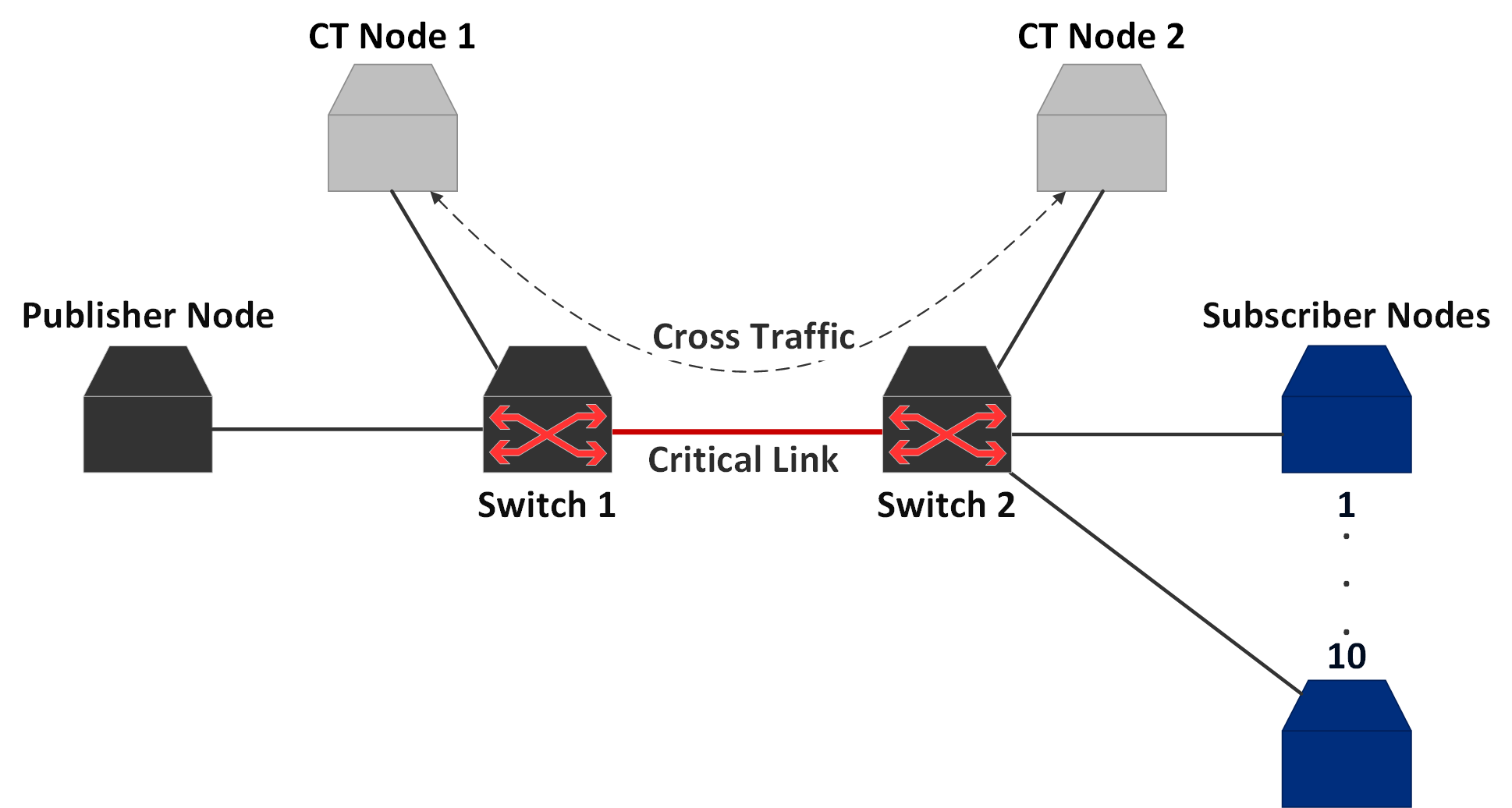}
    \caption{(A) Simple network consisting of a publisher, varying subscribers, and cross-traffic.}
    \label{fig:simplenetwork}
     \vspace{-3mm}
\end{figure}

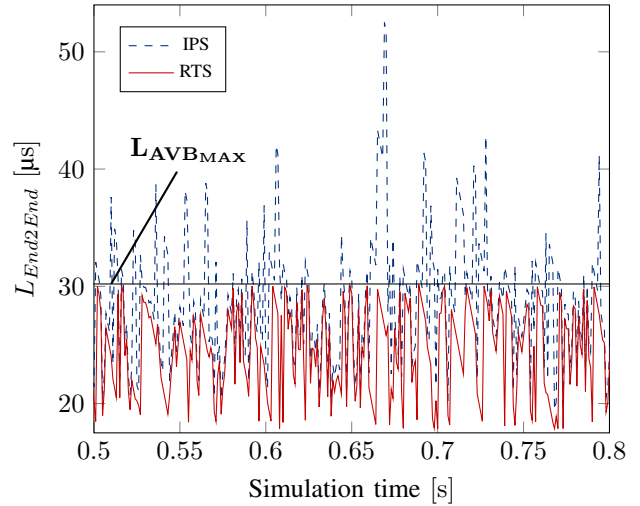
\begin{figure}
    \centering
        \begin{tikzpicture}

            \begin{axis}[
              change y base,
              y SI prefix=micro,
              y unit=\micro\second,
              x unit=\second,
              xlabel={Simulation time},
              ylabel={$L_{End2End}$},
              xmin=0.5,
              xmax=0.8,
              ymin=0.0000175,
              ymax=0.000054,
              legend style={font=\scriptsize,at={(.05,.95)},anchor=north west},
              ]


                \addplot [dashed, coreblue] table {\alltypeslatencytcp};
                \addplot [corered] table {\alltypeslatencyavb} ;
                \draw[black] (0,0.0000302) -- ( 2.5, 0.0000302);
                \node[coordinate, pin={[pin edge={black, thick}, pin distance=1.5cm]85:$\mathbf{L_{AVB_{MAX}}}$}] at (0.51, 0.0000302) { };
                \legend{\ac{IPS}, \ac{RTS}}
            \end{axis}
        \end{tikzpicture}
    \caption{End-to-end latency analysis with different service classes under varying \acl{CT}.}
    \label{fig:latency}
    \vspace{-3mm}
\end{figure}

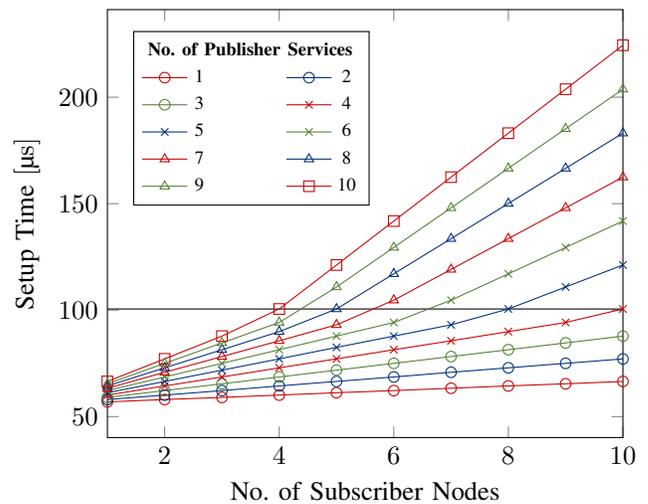
\begin{figure}
    \centering
        \begin{tikzpicture}[extended line/.style={shorten >=-#1,shorten <=-#1}, extended line/.default=1cm]

            \begin{axis}[
              change y base,
              y SI prefix=micro,
              y unit=\micro\second,
              xlabel={No. of Subscriber Nodes},
              ylabel={Setup Time},
              xmin=1,
              xmax=10,
              legend columns=2,
              legend style={font=\scriptsize,at={(.05,.95)},anchor=north west}]

                \draw [extended line=5cm] (5,0.000100536) -- (4,0.000100536);
                \addlegendimage{empty legend}
                \addlegendimage{empty legend}
                \addplot [mark=o,corered] table {\pubsubcountone} ;
                \addplot [mark=o,coreblue] table {\pubsubcounttwo} ;
                \addplot [mark=o,coregreen] table {\pubsubcountthree} ;
                \addplot [mark=x,corered] table {\pubsubcountfour} ;
                \addplot [mark=x,coreblue] table {\pubsubcountfive} ;
                \addplot [mark=x,coregreen] table {\pubsubcountsix} ;
                \addplot [mark=triangle,corered] table {\pubsubcountseven} ;
                \addplot [mark=triangle,coreblue] table {\pubsubcounteight} ;
                \addplot [mark=triangle,coregreen] table {\pubsubcountnine} ;
                \addplot [mark=square,corered] table {\pubsubcountten};

                \addlegendentry{\hspace{-.6cm}\textbf{No. of Publisher}}
                \addlegendentry{\hspace{-.7cm}\textbf{Services}}
                \addlegendentry{\hspace{-1cm}1}
                \addlegendentry{2}
                \addlegendentry{\hspace{-1cm}3}
                \addlegendentry{4}
                \addlegendentry{\hspace{-1cm}5}
                \addlegendentry{6}
                \addlegendentry{\hspace{-1cm}7}
                \addlegendentry{8}
                \addlegendentry{\hspace{-1cm}9}
                \addlegendentry{10}
            \end{axis}
        \end{tikzpicture}
    \caption{Setup times for increasing numbers of subscriber nodes. In each run a new publisher service is added to which all subscribers subscribe.}
    \label{fig:pubsubcount}
     \vspace{-3mm}
\end{figure}

(A)~The simple network visualized in Figure \ref{fig:simplenetwork}. It consists of a node that hosts several publishers, several subscriber nodes, two nodes that generate cross traffic, and two switches.
To evaluate different configurations, we will vary the number of publishers on the publisher node, the number of subscriber nodes, and the number of subscribers on each subscriber node.

(B)~A realistic automotive network based on a real communication matrix.
This database contains anonymized information for the sender, receiver and timings of all CAN messages in a production vehicle.

\subsection{Simple Network Evaluation}
Based on the simple network of Figure \ref{fig:simplenetwork} we measure setup times and latencies for different \ac{QoS} classes.
To observe the required time of each \ac{QoS} negotiation separately, the first set of simulation runs investigates
one subscriber using different \ac{QoS} classes without \ac{CT}.
Essentially, the setup time of a service consists of the time for \ac{QoS} negotiation plus the time to establish connectivity between a publisher and subscriber.
The duration of the \ac{QoS} negotiation lasts about \SI{76}{\micro\second} independent of the \ac{QoS} class.
For \ac{IPS}, a TCP connection has been established after \SI{130}{\micro\second} and UDP is ready after \SI{60}{\micro\second}.
For \ac{RTS} using AVB the connection has been established after \SI{100}{\micro\second},
while connectionless services are available right after the negotiation.
The different times for TCP and AVB result from different overheads for the connection establishment.

Figure \ref{fig:latency} shows the latency behavior of \ac{RTS} and \ac{IPS} with \ac{CT}.
$L_{\mbox{\small AVB}_{max}}$ is the latency limit \ac{AVB} requires for this topology.
The left \ac{CT}-Host sends full size frames in a normal distribution between $\approx$~\SI{2}{\micro\second} and \SI{23}{\micro\second} (resulting in a bandwidth $\approx$~\SI{950}{\mega\bit\per\second}).
The hardware delay of the switches is \SI{8}{\micro\second} and the processing time of the publisher and subscribers \SI{20}{\nano\second} each.
Since only one link with \ac{CT} exists in our simple network, the maximal latency must not exceed \SI{30.2}{\micro\second}.
This configuration consists of three subscribers, one of which using \ac{RTS} and the other two use \ac{IPS} as QoS-class.
In the simulation a maximum latency of \SI{93}{\micro\second} for \ac{IPS} and \SI{30}{\micro\second} for \ac{RTS} was captured in a \SI{20}{\second} run.
This shows that mixing different service classes does not exceed $L_{\mbox{\small AVB}_{max}}$ for \ac{RTS} traffic.

Figure \ref{fig:pubsubcount} compares the setup times as functions of the number of subscriber nodes for different publisher service multiplicities.
All publisher services run on the same node, while for each publisher service one subscriber service runs on each subscriber node.
The figure shows that the setup times are increasing (over-)proportionally with the number of QoS negotiations in the network.
The horizontal line in the figure at the \SI{100}{\micro\second} mark for the setup time indicates a change in the linear behavior at approximately 40 negotiations.
The network becomes overloaded at this point and  traffic jams delay the negotiation process.
Since the services of a car will be started stepwise, this situation should not occur in a real car network.

\begin{figure}
    \centering
    \includegraphics[width=\linewidth, trim= 0.1cm 0.0cm .95cm 2.7cm, clip=true]{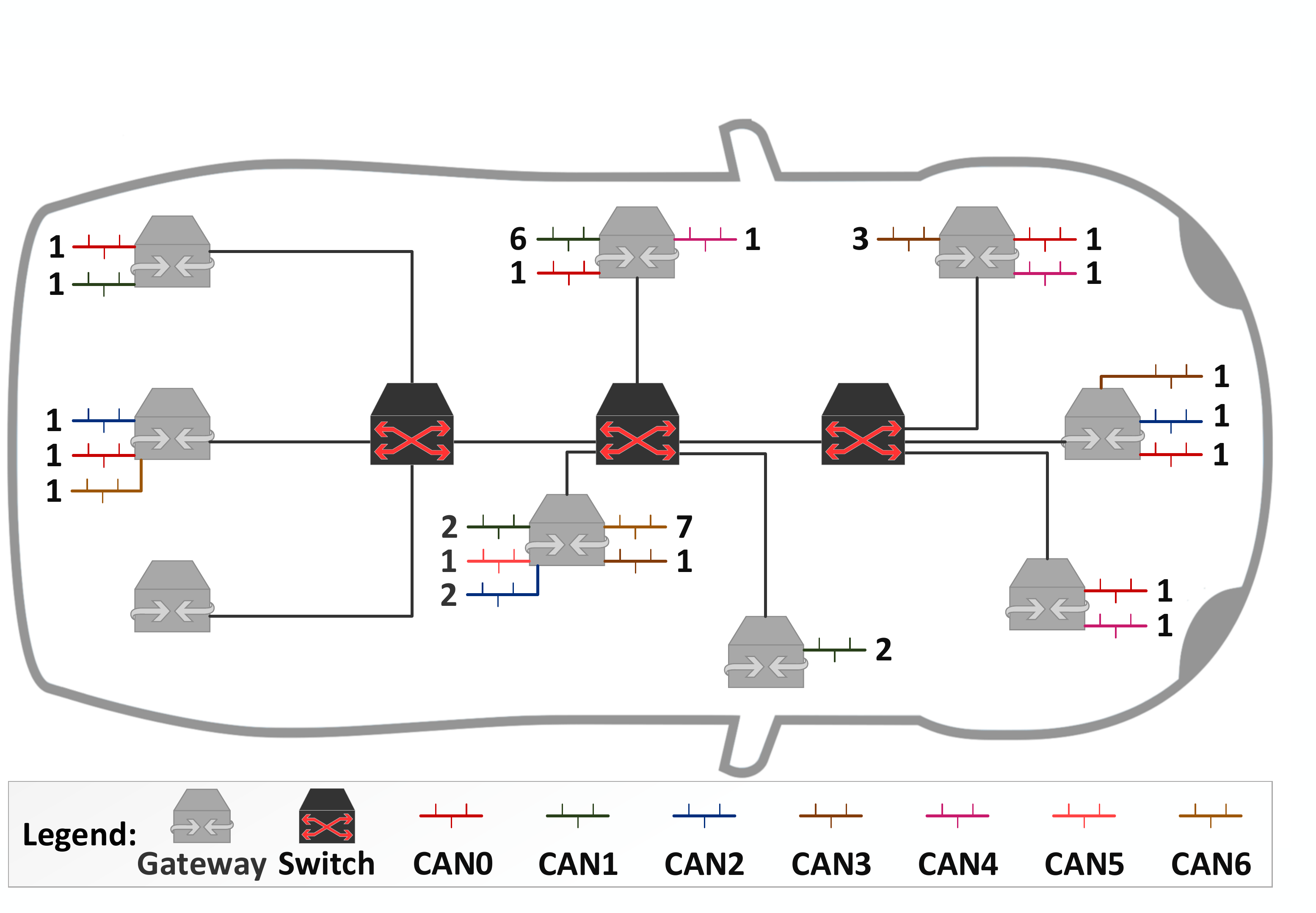}
    \caption{(B) Realistic automotive network based on real in-car communications.}
    \label{fig:recbarnetwork}
\end{figure}

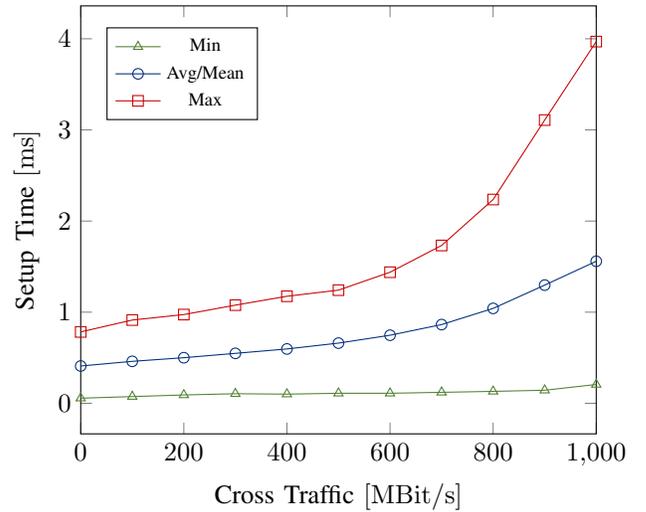
\begin{figure}
    \centering
        \begin{tikzpicture}

            \begin{axis}[change y base, y SI prefix=milli, y unit=\milli\second, x unit=\mega Bit\per\second,xlabel={Cross Traffic}, ylabel={Setup Time}, xmin=0, xmax=1000, legend style={font=\scriptsize,at={(.05,.95)},anchor=north west}]

                \addplot [mark=triangle,coregreen] table {\recbarmin} ;
                \addplot [mark=o,coreblue] table {\recbaravg} ;
                \addplot [mark=square,corered] table {\recbarmax} ;

                \legend{Min, Avg/Mean, Max}
            \end{axis}
        \end{tikzpicture}
    \caption{Minimum, maximum, and average setup time in a realistic network with \ac{CT}.}
    \label{fig:recbarctminmax}
\end{figure}

\subsection{Realistic Automotive Network Evaluation}
We now take a closer look at the realistic automotive network as visualized in
Figure \ref{fig:recbarnetwork}.
This network is a variation of a legacy network of an upper middle class production car.
The legacy network consists of domain-specific CAN-buses connected via a central gateway.
In the legacy network every \ac{ECU} of the same domain is connected to the CAN-bus of this domain regardless of its location in the car.
In our variation these CAN-buses are split up into sub-buses according to nine spatial zones distributed across the vehicle.
A CAN-to-Ethernet gateway connects each sub-bus to a switched Ethernet backbone.
Within Figure \ref{fig:recbarnetwork}, each domain-specific CAN bus is represented by its own color.
The numbers marking the outer edges of each sub-bus denote the number of CAN nodes connected to this sub-bus.

A Corresponding gateway receives the CAN messages an ECU generates.
The gateway then acts as a publisher of this information and provides a service for each CAN message.
All gateways connected to CAN ECUs needing the information of such a CAN message will subscribe to those services.
If a subscribing service at a gateway receives Ethernet messages, it forwards them to the correct CAN-bus.

Figure \ref{fig:recbarctminmax} depicts the growths of the setup times with increasing \ac{CT}.
All links in the network will be loaded with \ac{CT} from \SI{0}{\mega\bit} until \SI{1000}{\mega\bit}.
For every simulation, we captured the minimum, average, and maximum setup time.
There are only slight changes in the minimum times, while the average and maximum times seem to rise almost exponentially.
At a link load of about \SI{60}{\percent} \ac{CT}, the increase in setup time is drastic enough that negotiations might not finish in time.
High volumes of \ac{CT} above a certain threshold will lead to packet loss, which we did not account for in our simulations.

However, the results of Figure \ref{fig:recbarctminmax} show protocol delays of a few milliseconds. These results clearly comply with the requirements of approximately \SI{150}{ms} to \SI{200}{ms} as discussed at the beginning of this section.



\section{Conclusion and Future Work} 
\label{sec:conclusion_&_future_work}

In this paper, we presented a service-oriented communication middleware tailored to the requirements of the car.
This approach can meet heterogeneous requirements using a variable protocol stack and a \acl{QoSNP}, which allows services to dynamically negotiate their demands.
This way \ac{QoS} guarantees can be provided to clients based on their \ac{QoS} class.
We evaluated our concept following a case study in simulations based on a realistic automotive network.
We could show that the middleware and its dynamic \ac{QoS} negotiations successfully support mixed communication networks of heterogeneous requirements.
By using a multi-protocol stack, variable \ac{QoS} regimes can interoperate while preserving the flexibility of service-oriented communication.
We analyzed the impact of the negotiation on the setup time of the network. Our findings indicate that up to  70~\% cross-traffic are compliant with a maximum  of \SI{2}{\milli\second} for the setup, which is clearly acceptable for most of the traffic of the automotive network.
For safety-critical traffic that is particularly susceptible to jitter, we recommend the use of statically defined \ac{TDMA} classes, which experience no delay.

In future work, we will build a demonstrator of a real car components to determine real-world runtime delays and to analyze the various interactions within such a system.




\bibliographystyle{IEEEtran}
\bibliography{bib/bibliography,HTML-Export/all_generated}

\end{document}